\documentclass[a4paper,11pt]{article}
\usepackage[normalem]{ulem}
\usepackage{overpic,comment,booktabs}

\usepackage{pos}
\usepackage{soul}
\usepackage{xcolor}

\title{Electromagnetic form factors and structure of the $T_{bb}$ tetraquark}

\author*[a,b]{Ivan Vujmilovic}
\author[c]{Sara Collins}
\author[a,b]{Luka Leskovec}
\author[a,b]{Sasa Prelovsek}

\affiliation[a]{Jožef Stefan Institute, Jamova 39, 1000 Ljubljana, Slovenia}
\affiliation[b]{Faculty of Mathematics and Physics, University of Ljubljana, Jadranska ulica 19, 1000 Ljubljana, Slovenia}

\affiliation[c]{Institut f\"ur Theoretische Physik, Universit\"at Regensburg, 93040 Regensburg, Germany}

\emailAdd{ivan.vujmilovic@ijs.si}
\emailAdd{sara.collins@ur.de}
\emailAdd{luka.leskovec@ijs.si}
\emailAdd{sasa.prelovsek@ijs.si}

\abstract{We present the first lattice QCD calculation of electromagnetic form factors of a tetraquark, focusing on  the  $T_{bb} = bb\bar u \bar d$ with quantum numbers $I(J^P) = 0(1^+)$. The electromagnetic current probes the charge monopole, magnetic dipole and the electric quadrupole distributions within the tetraquark. From it, we find evidence that its structure consists of a compact heavy diquark $[bb]$ in spin one, color-antitriplet configuration, and a light antidiquark $[\bar u \bar d]$ in spin zero, color-triplet configuration. The computations were performed on a single CLS ensemble with $N_f = 2+1$ dynamical quarks at a lattice spacing $a\approx 0.064$ fm and with a pion mass $m_\pi \approx 290$ MeV.} 

\FullConference{The 42nd International Symposium on Lattice Field Theory (LATTICE2025)\\
2-8 November 2025\\
Tata Institute of Fundamental Research, Mumbai, India\\}

\begin{document}
\maketitle

\section{Introduction}\label{sec:intro}
During the past two-and-a-half decades we have witnessed an ever-growing number of discoveries of hadronic resonances in experiment \cite{ref:x3872,Belle:2007hrb,BESIII:2013ris,
BaBar:2005hhc,LHCb:2015yax,LHCb:2019kea,CDF:2009jgo,LHCb:2016axx,LHCb:2021vvq}, whose properties are inconsistent with conventional meson ($\bar q_1 q_2$) or baryon ($q_1 q_2 q_3$) valence quark content. These exotic states,  part of the QCD spectrum, naturally fall into the categories of tetraquarks ($\bar q_1 \bar q_2 q_3 q_4$), pentaquarks ($\bar q_1 q_2 q_3 q_4 q_5$), hybrid mesons ($\bar q_1 g q_2$) or glueball-like mesons. \\
\indent Currently there are several yet-undiscovered exotic tetraquark candidates from model calculations and lattice studies that are widely considered to be strongly stable. The most promising among those is the doubly-bottom tetraquark $T_{bb}$, in the $I(J^P) = 0(1^+)$ channel \cite{Bicudo:2012qt,Hudspith:2023loy,Leskovec:2019ioa,Colquhoun:2024jzh,
Alexandrou:2024iwi,Aoki:2022xxq,Junnarkar:2018twb,Prelovsek:2025vbr,Tripathy:2025vao}. The majority of studies find its mass to be $\mathcal{O}(100 \ \mathrm{MeV})$ below the $BB^*$ threshold. However, its detection in exclusive decays seems unlikely, while there may be a possibility of observing $T_{bb}$ in inclusive decays \cite{Gershon:2018gda}.\\
\indent The $T_{bb}$ tetraquark naturally lends itself to lattice QCD investigation of its structure. Since it is stable with respect to strong decay, volume effects in matrix elements are exponentially suppressed. It also presents an attractive opportunity to discern how the \textit{diquark-antidiquark} and molecular \textit{meson-meson} binding mechanisms manifest themselves in observables related to structure. The diquark-antidiquark structure of the $T_{bb}$ would be characterized by a compact diquark $[bb]_{\bar c}$ and a light antidiquark $[\bar u \bar d]_{c}$, where $c=3, \bar 6$ are available color configurations. A possible definition of a meson-meson structure implies a pair of spatially separated interacting quark-antiquark pairs $(\bar u b)_{c}(\bar d b)_{c}$ with $c=1, 8$, and spin correlations among quarks in each meson. Most quark model calculations (see e.g. \cite{Jaffe:2004ph,Vijande:2009kj}) combined with the large binding energy of the $T_{bb}$ observed in lattice studies favor the compact diquark-antidiquark $[bb]_{\bar 3}[\bar u \bar d]_3$ configuration. In our main work \cite{Vujmilovic:2025czt} and here we present the first lattice QCD calculation of electromagnetic form factors of the $T_{bb}$. In addition to this, we extract the charge form factors of $B$ and $B^*$ mesons to assess the relative compactness of the exotic tetraquark in comparison to the combined size of $B$ and $B^*$, which represent the closest decay threshold.
\section{$\langle h(p_2, \lambda_2) | \hat{\jmath}^{\mu}_{EM} | h(p_1, \lambda_1) \rangle$ form factor decomposition}\label{sec:ff}
Studying the composition of hadrons in terms of their elastic electromagnetic (EM) form factors offers the advantage of having a clear physical interpretation. These form factors are encoded in EM matrix elements, defined in the infinite-volume spacetime continuum as
\begin{gather}
    \mathcal{M}_{EM}^\mu = \langle h(p_2,\lambda_2)  | \hat{\jmath}^\mu_{EM, cont}(x=0)| h(p_1, \lambda_1)\rangle,  \hspace{1cm}
    \hat{\jmath}^\mu_{EM,cont} \equiv \sum_{q} e_q \bar q \gamma^\mu q ,\label{eq:eq1}
\end{gather}
where $e_q$ is the electric charge of quark $q$ and $h(p_{1(2)},\lambda_{1(2)})$ is the ingoing(outgoing) state $h$ with momentum $p_{1(2)}$ and helicity $\lambda_{1(2)}$. The vector current and consequently the matrix element \eqref{eq:eq1} is additive with respect to quark flavors, allowing one to meaningfully decompose form factors into sums $F(Q^2) = \sum_{q}F_q(Q^2)$ and probe the distributions of individual quarks inside a hadron. Lorentz-covariant parametrization of eq.~\eqref{eq:eq1} for (pseudo)scalars gives a single charge form factor $F_C$
\begin{gather}
    \mathcal{M}_{EM}^\mu = (p_1 + p_2)^\mu F_C(Q^2),
\end{gather}
while for $J^P = 1^\pm$ hadrons we get
\begin{align}\label{eq:eq2}
    \mathcal{M}_{EM}^\mu = -(&p_1 +  p_2)^\mu  (\varepsilon_2^*  \cdot \varepsilon_1  ) F_1 (Q^2 ) 
    - [(\varepsilon_2^*  \cdot q)\varepsilon_1^\mu - (\varepsilon_1  \cdot q)\varepsilon_2^{*\mu}] F_2(Q^2) + \nonumber \\[5pt] &+ \frac{(\varepsilon_2^* \cdot q) (\varepsilon_1 \cdot q)}{2m^2} (p_1 + p_2)^\mu F_3(Q^2) ,
\end{align}
with $Q^2 \equiv -(p_2-p_1)^2 > 0$ and $\varepsilon^{(*)}_{1(2)}$ being momentum transfer and polarization four-vectors, respectively. The form factors in eq.~\eqref{eq:eq2} are related to charge, magnetic dipole and electric quadrupole form factors via a linear transformation
\begin{align}\label{eq:eq3}
    \begin{pmatrix}
    F_C (Q^2) \\[1pt]
    F_M (Q^2) \\[1pt]
    F_{\cal{Q}} (Q^2)
    \end{pmatrix}
    =
    \begin{pmatrix}
    1+\frac{2}{3}\eta & - \frac{2}{3}\eta & \frac{2}{3}\eta(1+\eta) \\[1pt]
    0 & 1 & 0 \\[1pt]
    1 & -1 & (1 + \eta)
    \end{pmatrix}
    \begin{pmatrix}
    F_1 (Q^2) \\[1pt]
    F_2 (Q^2) \\[1pt]
    F_3 (Q^2)
    \end{pmatrix} ,
\end{align}
with $\eta = \frac{Q^2}{4m^2}$. In the infinite-volume spacetime continuum the elastic charge form factors are normalized to the total electric charge $Z$ of the state in units of elementary charge $e_0$ at $Q^2=0$: $F_C(0) \equiv Z$, while the electric quadrupole and magnetic dipole form factors yield the corresponding values of multipole moments: $\cal Q $ $= \frac{1}{m^2} F_{\cal Q}(0)$ and $\mu = \frac{1}{2m} F_M(0)$, respectively. 
\section{Lattice setup}
We employed a single ensemble of gauge configurations (denoted \texttt{X253}) with a hypervolume $N_L^3 \times N_T = 40^3 \times 128$ and lattice spacing $a=0.06379(37)$ fm, generated by the Coordinated Lattice Simulations (CLS) consortium \cite{Bruno:2014jqa}. The ensemble features $N_f = 2+1$ dynamical quarks, with degenerate $u/d$ quarks. They are described by a non-perturbatively $O(a)$ improved Wilson action, resulting in a pion mass $m_\pi \approx 290$ MeV. The $b$ quarks are implemented with an anisotropic relativistic heavy quark action \cite{El-Khadra:1996wdx,Chen:2000ej}, tuned to yield the physical masses and the correct continuum energy-momentum dispersion relations of $B$ and $B^*$ mesons. More details on the heavy quark action tuning are given in the main work and its accompanying Supplemental Material \cite{Vujmilovic:2025czt}.\\
\indent The light and heavy EM currents on the lattice are renormalized with factors $Z^V_{u/d}$ and $Z^V_b$ and added to give the total current $\hat{\jmath}^\mu_{EM}$
\begin{gather}
\hat \jmath_{u/d}^\mu= Z_{u/d}^V \cdot \left( \tfrac{2}{3}\bar u\gamma^\mu u-\tfrac{1}{3}\bar d \gamma^\mu d\right), \hspace{0.3cm} \hat\jmath_{b}^\mu=Z_b^V \cdot \left( -\tfrac{1}{3}\bar b  \gamma^\mu b \right)  \Longrightarrow \hat\jmath_{EM}^\mu=\hat\jmath_{  u/d}^\mu+\hat\jmath_{ b}^\mu.  \label{eq:jEM}
\end{gather}
The renormalization conditions, fixing the values of $Z^V_{u/d}$ and $Z^V_{b}$ are given by
\begin{gather}
    \langle T_{bb}(\vec p = \vec 0) | \hat{\jmath}^{0}_{u/d} | T_{bb}(\vec p = \vec 0) \rangle \equiv -\frac{1}{3}, \hspace{1cm} \langle T_{bb}(\vec p = \vec 0) | \hat{\jmath}_b^0 | T_{bb}(\vec p = \vec 0) \rangle \equiv -\frac{2}{3} .
\end{gather}
In this study $s-$ and $c-$quark currents are omitted, since they result in disconnected contributions to the total three-point correlation functions, which were not calculated in this study.\\
\begin{figure}[t]
    \centering
    \begin{overpic}[width=0.36\columnwidth]{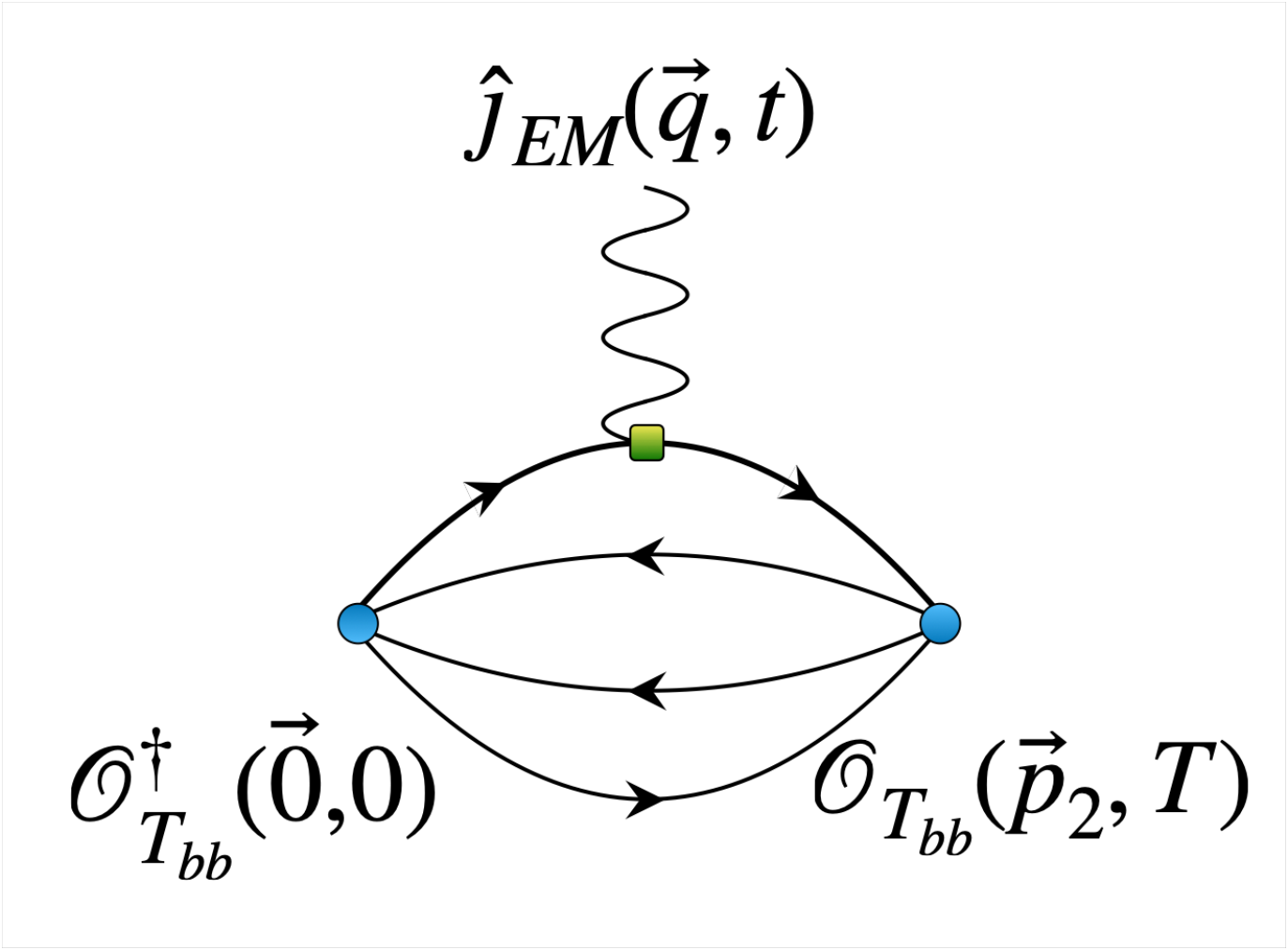}
    \put(0,71){}
    \end{overpic}
    \caption{Generic connected diagram generated by the Wick contractions in the $T_{bb}$ three-point correlator \eqref{eq:3ptformula}.}
    \label{fig:threept}
\end{figure}
\indent To extract the desired EM matrix elements of hadrons $h = T_{bb}, B, B^*, \pi$, we computed the corresponding two- and three-point correlation functions
\begin{align}
    \mathcal{C}_2 & ( \vec p, t)  = \langle \Omega | \mathcal{O}_{h} (\vec p, t) \mathcal{O}_h^\dagger (x=0) | \Omega \rangle, \label{eq:2ptformula} \\[8pt]
    \mathcal{C}_{3}^\mu &(\vec p_2 = \vec 0, \vec q, T; t) = \langle \Omega | \mathcal{O}_{h} ( \vec p_2 = \vec 0, T) \hat \jmath_{EM}^\mu (\vec q, t) \mathcal{O}^\dagger_{h} (  x=0) | \Omega \rangle = \nonumber \\
    &= \sum_{n,m = 0}^\infty \frac{\mathcal{Z}^{f*}_n \mathcal{Z}^i_m}{(2E^f_n)(2E^i_m)}  \mathcal{M}_{nm}^\mu \ e^{-E^f_n (T-t)} e^{-E^i_mt}, \label{eq:3ptformula}
\end{align}
where $\mathcal{O}_h$ are interpolators that couple to hadrons $h$ with definite momenta as indicated in the brackets, inserted at times $0$ at the source and $\frac{T}{a}=12,15,18,22$ at the sink in three-point correlators. The EM current is inserted at all intermediate times $t \in [0, T]$. The interpolators generate nonzero overlap $\mathcal{Z}^{i(f*)}_{n(m)}$ to an infinite tower of states with energies $E^{i(f)}_{n(m)}$ at the source~($i$) and the sink~($f$), and are summed over by indices $n,m$ in eq.~\eqref{eq:3ptformula}. Isolating the matrix element $\mathcal{M}_{00}^\mu$ requires knowing the source(sink) ground-state overlap factors $\mathcal{Z}_0^{i(f*)}$ and energies $E_0^{i(f)}$, obtained from fits to the two-point correlators \eqref{eq:2ptformula}. A single interpolator was used for each hadron $h$ with more details provided in \cite{Vujmilovic:2025czt}. Furthermore, the interpolators in eqs.~\eqref{eq:2ptformula} and \eqref{eq:3ptformula} are projected to definite irreps of the cubic (sub)groups; we omit labels indicating the irreps for brevity.\\
\indent The ratio $R_3^\mu$, defined as
\begin{align}
    R_3^\mu (\vec p_2 = \vec 0, \vec q, T; t) = \frac{(2E_0^f) (2E_0^i)}{\mathcal{Z}_0^{f*}\mathcal{Z}_0^i} e^{E_0^f(T-t)} e^{E_0^i t} \ \mathcal{C}_3^\mu (\vec p_2 = \vec 0, \vec q, T; t), \label{eq:ME}
\end{align}
was employed in extracting matrix elements. In the limit of large Euclidean time separations, only $\mathcal{M}_{00}^\mu$ appears in eq.~\eqref{eq:ME} with subleading contributions from excited states. Two models were used in fitting $R_3^\mu$: a constant model $\mathcal{M}$, assuming zero excited state contamination and a model incorporating the first excited state, $\mathcal{M} + \alpha e^{-\Delta E_1 t} + \beta e^{- \Delta E_2 (T-t)}$. Prior to fitting, we averaged the ratio \eqref{eq:ME} over multiple equivalent directions $\hat q$ that yield the same value of $Q^2$. 
\section{Results}
\begin{figure*}[t]
    \begin{overpic}[width=0.325\textwidth]{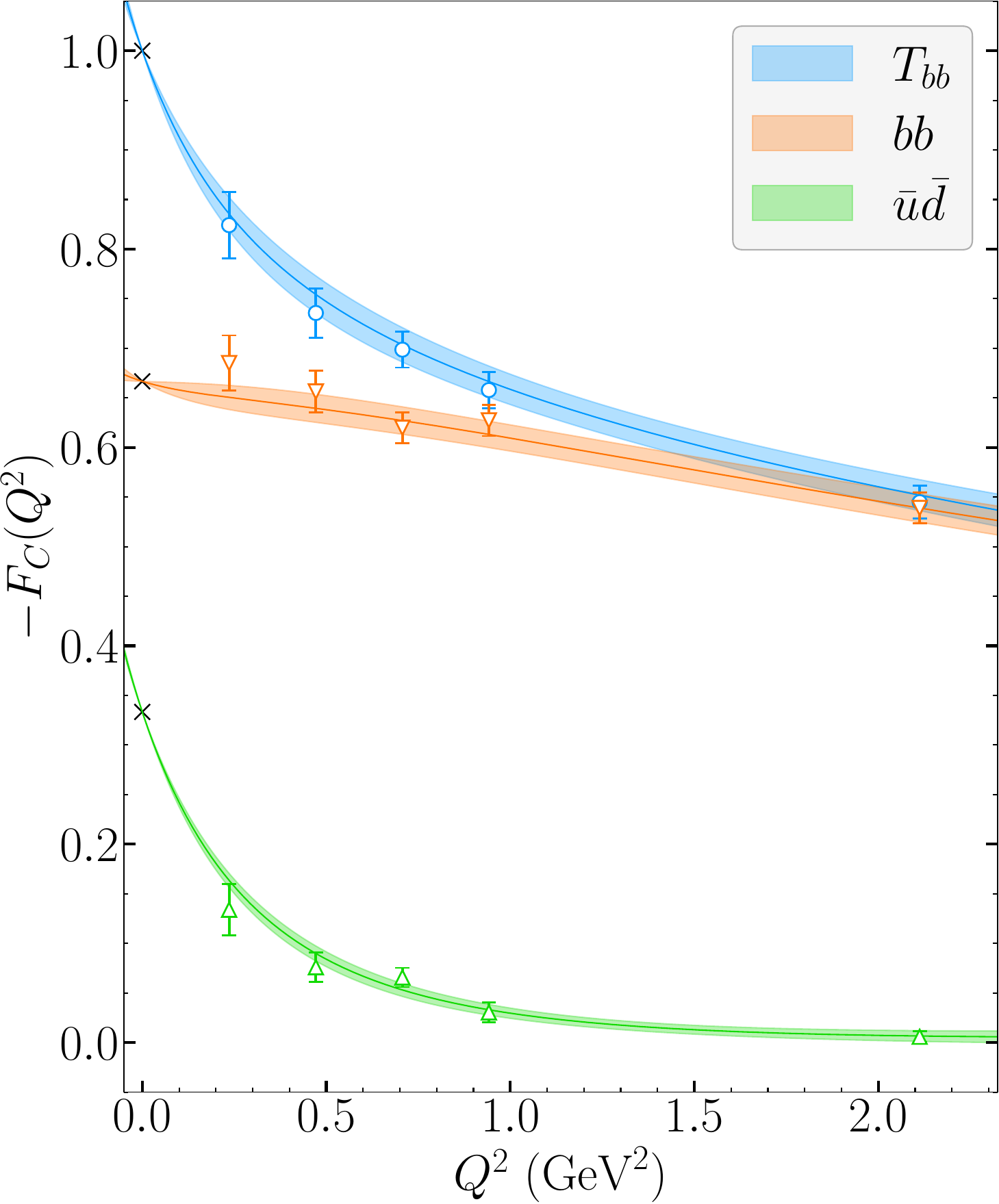}
    \put(0,99){\textbf{a)}}
    \end{overpic}
    \hfill
    \begin{overpic}[width=0.325\textwidth]{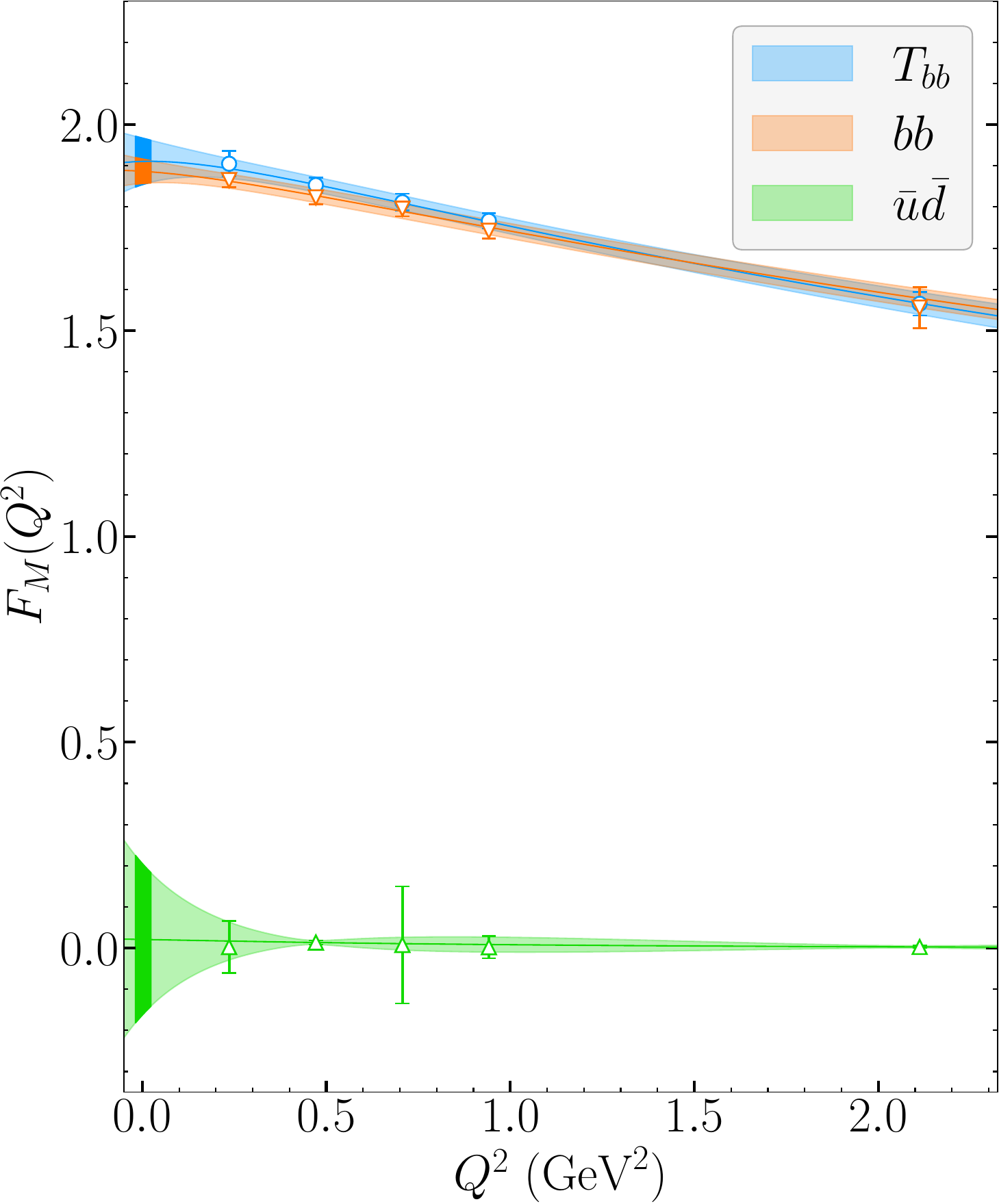}
    \put(0,99){\textbf{b)}}
    \end{overpic} 
    \hfill
    \begin{overpic}[width=0.325\textwidth]{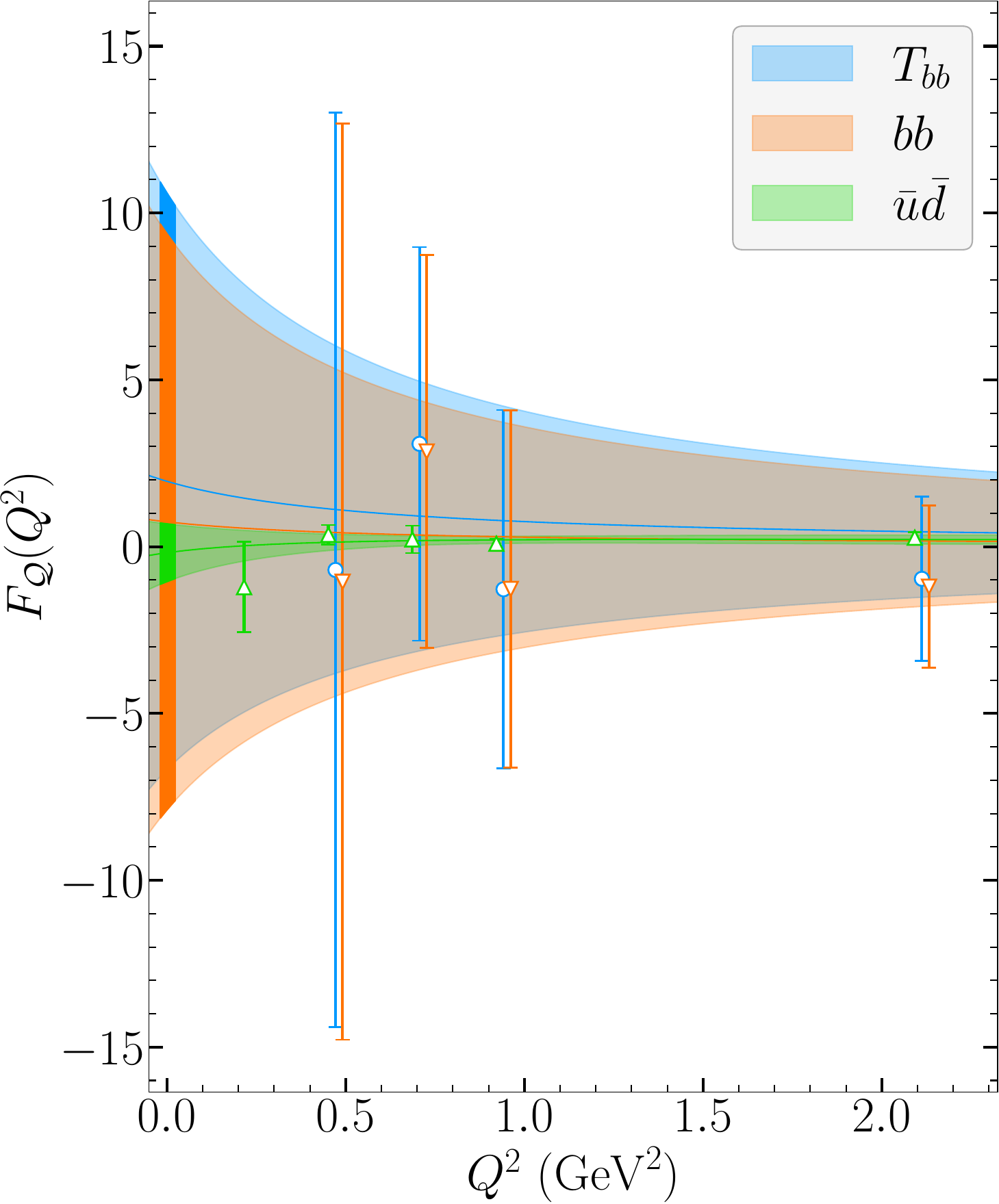}
    \put(0,99){\textbf{c)}}
    \end{overpic} 
    \caption{Form factors of the $T_{bb}$. Each subfigure shows the total value of the form factors with separate contributions yielded by the light current 
  $\hat \jmath_{u/d}^\mu$ and the heavy current $\hat \jmath_b^{\mu}$. The crosses in \textbf{a)} mark values to which each of the charge form factors have been normalized at $Q^2=0$, while the shaded vertical bands in \textbf{b)} and \textbf{c)} indicate the values of magnetic dipole moments $2m_{T_{bb}}\cdot \mu$ and electric quadrupole moments $m_{T_{bb}}^2 \cdot \cal Q$, respectively, also found in Table \ref{tab:tbb_moments}. The points on the rightmost plot are slightly horizontally displaced for improved visibility.}
    \label{fig:tbb_formfac}
\end{figure*}

The masses of $T_{bb}$, $B$ and $B^*$ are shown in Table \ref{tab:r2ch}, indicating a significant binding energy of the tetraquark
\begin{align}
    m_{T_{bb}} - (m_B + m_{B^*}) = -64(10) \ \mathrm{MeV},
\end{align}
consistent with other studies.\\
\begin{figure*}[t]
    \begin{overpic}[width=0.49\textwidth]{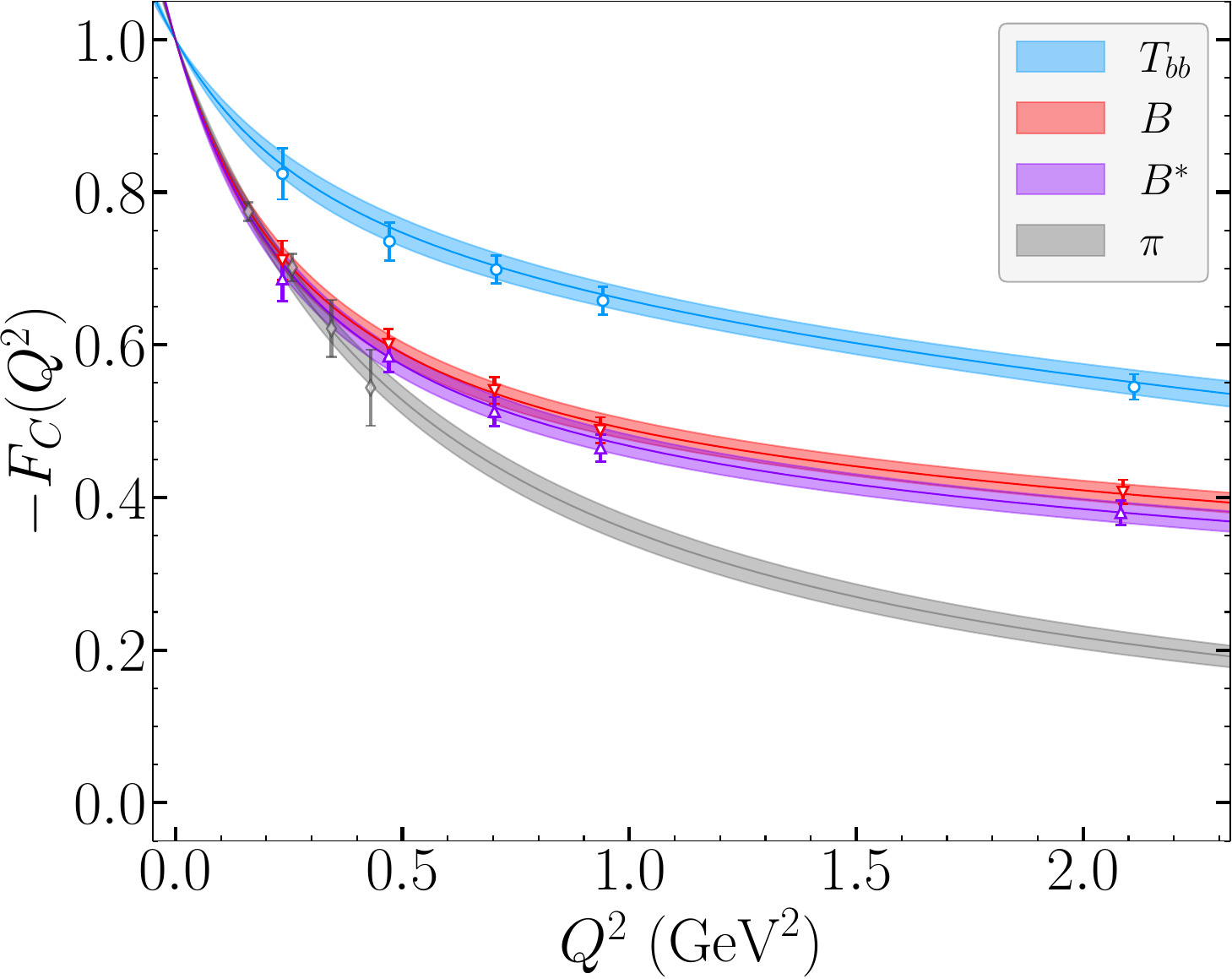}
        \put(0,79){\textbf{a)}}
    \end{overpic}
    \hfill
    \begin{overpic}[width=0.49\textwidth]{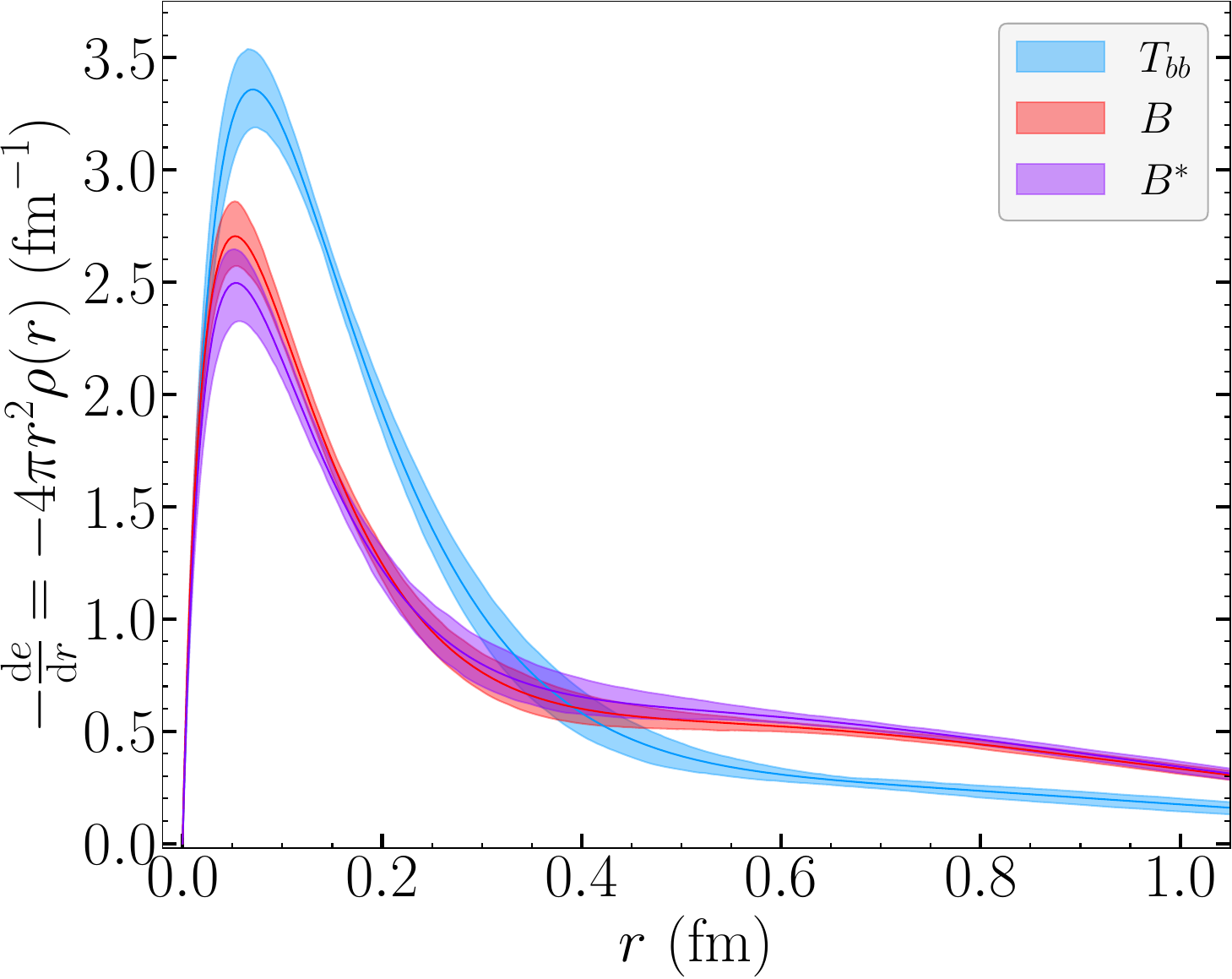}
        \put(0,79){\textbf{b)}}
    \end{overpic} 
  \caption{\textbf{a)} Charge form factors of $T_{bb}, B= b\bar u, B^* = b\bar u , \pi = d \bar u$, shown as a function of $Q^2$. Discrete points represent lattice data, while the continuous bands show $z-$expansion fits. Second order expansion (up to and including $n\!=\!2$ in eq.~\eqref{eq:zexp}) was used to parametrize $T_{bb}, \ B, \ B^*$ electric form factors, while a first order expansion sufficed for an adequate parametrization of the pion form factor. \textbf{b)} Position-space charge densities $\rho(r)$, represented in the form of $-\tfrac{\mathrm{d}e}{\mathrm{d}r}=-4\pi r^2\rho$, are related to the form factors via a Fourier transform. Negative values of charge form factors and distributions are shown, given that considered hadrons are negatively charged.}
  \label{fig:fc_comp}
\end{figure*}
\indent The main results of this work are shown in Figures \ref{fig:fc_comp} and \ref{fig:tbb_formfac}. Lattice data is represented by discrete markers, while the continuous $Q^2$ dependence of form factors $F$ is parametrized with the $z-$expansion featuring a pole term
\begin{align}
    F(Q^2) = \frac{1}{1 + \frac{Q^2}{m_r^2}} \sum_{n=0}^{N_{max}} a_n z^n (Q^2), \hspace{1cm} z(Q^2) = \frac{\sqrt{t_+ + Q^2} - \sqrt{t_+ - t_0}}{\sqrt{t_+ + Q^2} + \sqrt{t_+ - t_0}} , \label{eq:zexp} 
\end{align}
where the definition of the variable $z$ is given by the right equation. In eq.~\eqref{eq:zexp}, $a_n$ are expansion coefficients serving as fit parameters and $m_r$ is the resonance mass. Parameter $t_+$ is the value of the lowest multi-particle threshold in a given channel, while $t_0$ is a tunable parameter. Resonances $r$ and thresholds $t_+$ appearing in eq.~\eqref{eq:zexp} depend on the quantum numbers of the hadron that is analyzed and are summarized in Table \ref{tab:r2ch}. Throughout all fits we employ the lattice pion mass $m_\pi \approx 290$ MeV and PDG values of $\rho/\omega$ masses since they are not known for our ensemble. Nonetheless, we have verified that fits remain robust when varying the masses within a reasonable range.\\
\indent Figure \ref{fig:fc_comp}\textbf{a} shows the charge form factors of all four hadrons of interest. The charge radius $\sqrt{\langle r_C^2 \rangle} \equiv \sqrt{6 \cdot \frac{\mathrm{d}F_C}{\mathrm{d}Q^2}(0)}$ of the $T_{bb}$ is found to be significantly smaller than the sum of radii of $B$ and $B^*$ mesons. This result favors diquark-antidiquark structure, as opposed to a molecular meson-meson binding; if the latter were the case, one would expect the $T_{bb}$ radius to be commensurate with the combined value of that for $B$ and $B^*$, as discussed in Section \ref{sec:intro}. The compactness of the $T_{bb}$ in comparison to the sizes of $B$ and $B^*$ mesons can also be established by calculating the Fourier transforms of their charge form factors with respect to momentum $\vec q$, giving the charge distributions in position-space, as shown in Fig.~\ref{fig:fc_comp}\textbf{b}. Fourier transform involves the integral over three-momenta $|\vec q|^2 \approx Q^2$; these are approximately non-relativistic in the energy range available for all hadrons considered except the pion, which is therefore omitted.\\
\indent Figure \ref{fig:tbb_formfac} shows the three form factors of $T_{bb}$. As discussed in Section \ref{sec:ff}, total values, as well as the individual contributions from the $u/d$ and $b$ quark currents to each of the form factors are determined. These are extracted from the matrix elements $\langle T_{bb} | \hat{\jmath}^\mu_{u/d} |T_{bb} \rangle$ and $\langle T_{bb} | \hat{\jmath}^\mu_b | T_{bb} \rangle$, respectively. The charge radius of the $[bb]$ diquark is significantly smaller than the charge radius of the light antidiquark $[\bar u \bar d]$, further favoring the diquark-antidiquark component over the meson-meson structure, consistent with the previous argument establishing the compactness of the $T_{bb}$ compared to the sizes of the $B$ and $B^*
$ mesons. Motivated by this, we introduce the set of all diquark-antidiquark states in QCD Hilbert space as
\begin{align}
    \{ [bb]_{\bar{c}}^{l_{bb}, s_{bb}} [\bar u \bar d]_{c}^{I = 0, l_{\bar u \bar d}, s_{\bar u \bar d}} \}^{J^P = 1^+}_{l_{12}} . \label{eq:diq}
\end{align}
The (anti)diquarks in eq.~\eqref{eq:diq} have definite quantum numbers: the spins $s_{bb}, s_{\bar u \bar d}$, orbital angular momenta $l_{bb}, l_{\bar u \bar d}$ and $SU(3)$ color configurations $c$ (where $c = 3, \bar 6$ are possible). The relative orbital angular momentum between the diquark-antidiquark is denoted by $l_{12}$. The total $T_{bb}$ state is in principle a linear combination of all vectors \eqref{eq:diq} that are consistent with the quantum numbers $I(J^P) = 0(1^+)$ and satisfy the Pauli principle. The Pauli principle is applicable to both the heavy quark and the light quark pair (in the isospin limit) \cite{Vijande:2009kj}
\begin{align}
    (a)  \hspace{0.3cm} & (-1)^{s_{bb} + l_{bb} + c} = 1, \hspace{1.4cm} (b) \hspace{0.3cm} (-1)^{s_{\bar u \bar d} + l_{\bar u \bar d} + c} = -1 \nonumber \\
    & (-1)^{l_{bb} + l_{\bar u \bar d} + l_{12}} = 1 \xrightarrow{(a), (b)} (-1)^{s_{bb} + s_{\bar u \bar d} + l_{12}} = -1 . \label{eq:pauli}
\end{align}
The antisymmetry of both pairs of quarks and antiquarks yields the first line, while the positive parity of $T_{bb}$ implies the second line. \\
\begin{table}[t!]
\centering
\begin{minipage}{0.49\textwidth}
  \centering
  \setlength{\tabcolsep}{3pt}
\begin{tabular}{l l l r c}
    \toprule
    $h$ & $m_h$ (GeV) & $\sqrt{\langle r_C^2 \rangle}$ ($\mathrm{fm}$) & $r$ & $t_+$ \\
    \midrule
    $T_{bb}$ & $10.5765(98)$ & $0.499(31)$ & $\omega$ & $(3m_\pi)^2$   \\
     $B$ & $5.3020(17)$& $0.692(21)$ & $\rho$ & $(2m_\pi)^2$ \\
     $B^*$ & $5.3387(20)$& $0.698(23)$ & $\rho$ & $(2m_\pi)^2$ \\
     $\pi$ & $0.28953(97)$& $0.652(20)$ & $\rho$ & $(2m_\pi)^2$ \\
    \bottomrule
  \end{tabular}
  \caption{Hadron masses $m_h$ and charge radii $\sqrt{\langle r_C^2 \rangle}$. Columns $r$ and $t_+$ show the resonances and multi-particle thresholds, respectively, that appear in the $z-$expansions.}
  \label{tab:r2ch}
\end{minipage}
\hfill
\begin{minipage}{0.49\textwidth}
  \centering
  \setlength{\tabcolsep}{3pt}
  \begin{tabular}{c c c c}
    \toprule
        {} & $\sqrt{\langle r_C^2 \rangle}$ (fm) & $2m_{T_{bb}} \cdot \mu$ & $m_{T_{bb}}^2 \cdot \cal{Q}$ \\
    \midrule
    $T_{bb}$ & $0.499(31)$ & $1.912(57)$ & $1.9(8.7)$ \\
    $[bb]$ & $0.174(59)$ & $1.887(29)$ & $0.6(8.8)$   \\
     $[\bar u \bar d]$ & $0.511(14)$& $0.02(18)$ & $-0.18(86)$ \\ 
    \bottomrule
  \end{tabular}
  \caption{Charge radii, magnetic dipole $\mu$ and electric quadrupole moments ${\cal Q}$ of $T_{bb}$ and the constituent (anti)diquarks. The same values are also shown with shaded bands at $Q^2=0$ in Figs. \ref{fig:tbb_formfac}\textbf{b} and \ref{fig:tbb_formfac}\textbf{c}.}
  \label{tab:tbb_moments}
\end{minipage}

\end{table}

\indent In the following analysis we turn to the remaining form factors in Figs.~\ref{fig:tbb_formfac}\textbf{b} and \ref{fig:tbb_formfac}\textbf{c} to further narrow down the degrees of freedom introduced in eq.~\eqref{eq:diq}. From them, one can extract the total magnetic dipole and electric quadrupole momenta as well as the contributions of individual quarks. These quantities are obtained by taking the limiting values at $Q^2 = 0$ of the $z-$expansions \eqref{eq:zexp} and can be found in Table \ref{tab:tbb_moments}. Specifically, the electric quadrupole momenta of $T_{bb}$ and its constituent (anti)diquarks are all found to be consistent with zero. This is possible only if all orbital angular momenta in \eqref{eq:diq} are zero and the orbital wave function is in the $S-$wave ($l_{bb} = l_{\bar u \bar d} = l_{12} = 0$). The magnetic dipole distribution within the $T_{bb}$, shown in Fig.~\ref{fig:tbb_formfac}\textbf{b}, unambiguously signals that the vast majority of the magnetic dipole moment is generated by the heavy diquark $[bb]$, while the $[\bar u \bar d]$ contribution is consistent with zero. Given that the magnetic dipole moment operator couples both to the orbital and spin components of the $T_{bb}$
\begin{align}
    \mu = \langle T_{bb} | \sum_{q=u,d,b} \frac{e_q}{2m_q} (\hat l_q + g_q \hat s_q ) | T_{bb} \rangle \equiv \frac{F_M(0)}{2m_{T_{bb}}} ,
\end{align}
and the former is found to be zero, we conclude that the $[bb]$ diquark is in spin one, while the $[\bar u \bar d]$ antidiquark is in the spin zero configuration
\begin{align}
    s_{\bar u \bar d} = 0, \hspace{0.5cm} s_{bb} = 1  .
\end{align}
\indent Finally, combining eq.~\eqref{eq:pauli} with the deduced values of spin and orbital quantum numbers uniquely determines the color configuration of the diquark-antidiquark to be $c = 3$. Therefore, the information obtained from the electromagnetic form factors of the $T_{bb}$ suffices to completely constrain its wave function as
\begin{align}
    |T_{bb} \rangle  =\{ [bb]_{\bar{3}}^{l_{bb}=0, s_{bb}=1} [\bar u \bar d]_{3}^{I = 0, l_{\bar u \bar d}=0, s_{\bar u \bar d}=0} \}^{J^P = 1^+}_{l_{12}=0} ,
\end{align}
wherein any other combination would be in tension with our lattice data at the current level of precision.
\begin{figure}[t]
    \centering
    \begin{overpic}[width=0.3\columnwidth]{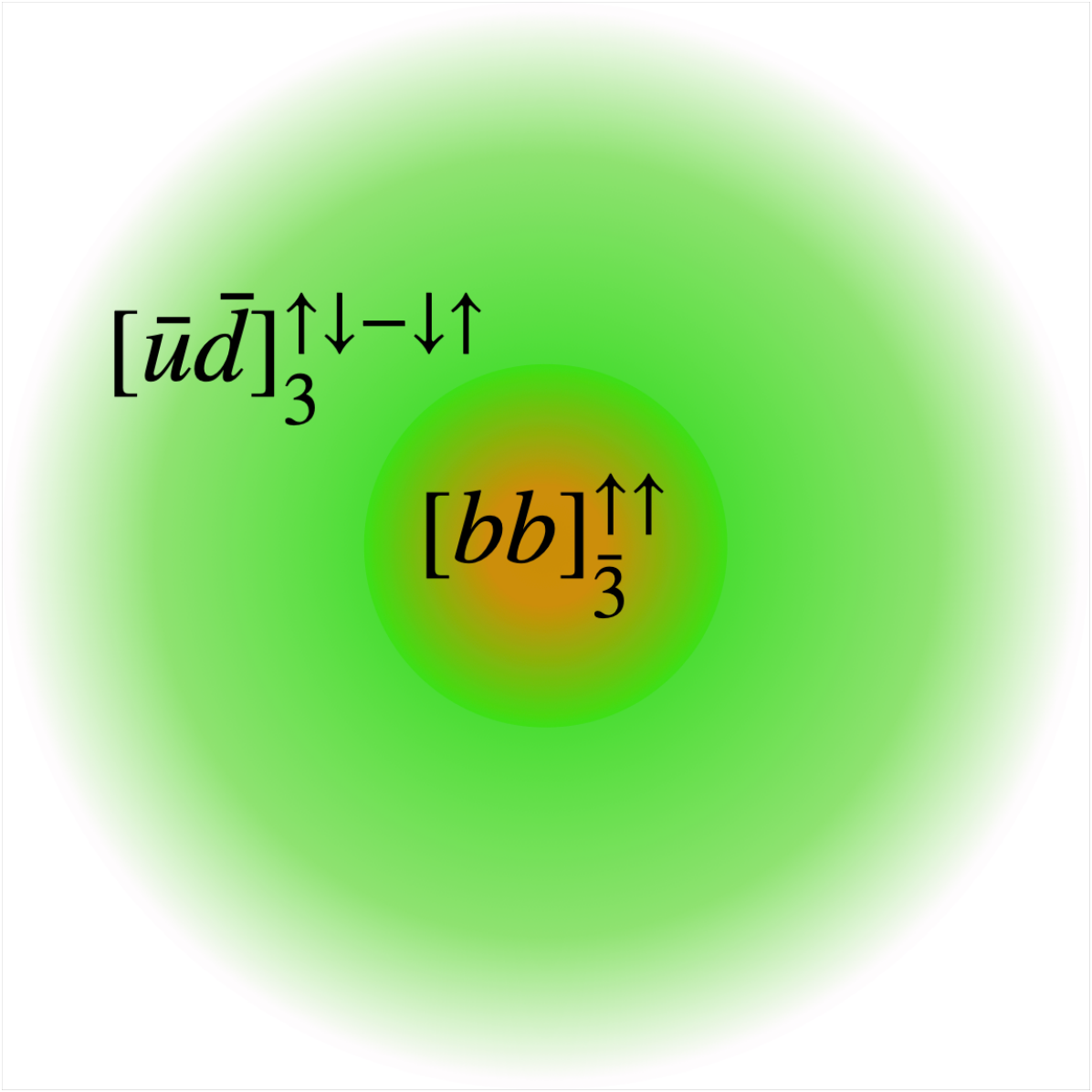}
    \put(0,71){}
    \end{overpic}
    \caption{Pictorial representation of the spatial, spin and color composition of the $T_{bb}$, as determined by from its EM form factors.}
    \label{fig:threept}
\end{figure}
\section{Summary and conclusions}
In this study, the first lattice QCD calculation and analysis of electromagnetic form factors of the $T_{bb}$ tetraquark was presented. The simulations were performed on a single CLS ensemble with an unphysical pion mass $m_\pi \approx 290$ MeV using $b-$quarks implemented with a relativistic heavy quark action tuned to reproduce the physical $B$ and $B^*$ masses and dispersion relations. Following this, we measured the binding energy of the $T_{bb}$ to be $m_{T_{bb}} - (m_B + m_{B^*}) = -64(10)$ MeV.\\
\indent Having established the existence of a $T_{bb}$ bound state in our lattice setup, we computed its matrix elements containing local heavy and light vector currents, enabling us to determine the separate contributions to the electromagnetic form factors from the light and heavy (anti)quarks. This approach allowed us to extract the charge distributions within the tetraquark, the magnetic dipole moment and the electric quadrupole moment, as well as their flavor decompositions. We found that the available lattice data, combined with the Pauli principle applied to heavy and light (anti)quark pairs is sufficient to completely describe the internal structure of the $T_{bb}$. Rather uniquely, it is composed as a bound state consisting of a compact heavy diquark $[bb]$ in a color-antitriplet with spin one, and a light antidiquark $[\bar u \bar d]$ in a color-triplet in spin zero.\\
\section*{Acknowledgements}
We thank M. Padmanath and R. J. Hudspith for valuable discussions, especially during the early stages of the project regarding the tuning of the $b-$quark action. We thank our colleagues who are part of CLS for their joint effort in the generation of gauge configurations that were employed in lattice simulations for this study. Chroma \cite{Edwards:2004sx} and QUDA \cite{Clark:2009wm} libraries were used in running lattice simulations. The authors gratefully acknowledge the HPC RIVR consortium (\href{www.hpc-rivr.si}{https://www.hpc-rivr.si}) and EuroHPC JU (\href{eurohpc-ju.europa.eu}{https://eurohpc-ju.europa.eu}) for funding this research by providing computing resources of the HPC system Vega at the Institute of Information Science (\href{www.izum.si}{https://www.izum.si/en/home}), in particular the project QCD on Vega (S24O01-37 and 525002-11). The authors also acknowledge the scientific support and HPC resources provided by the Erlangen National High Performance Computing Center (NHR@FAU) of the Friedrich-Alexander-Universität Erlangen-Nürnberg (FAU) under the NHR project b124da. NHR funding is provided by federal and Bavarian state authorities. NHR@FAU hardware is partially funded by the German Research Foundation (DFG) – 440719683.  The work of I. V., L. L. and S. P. is supported by the Slovenian Research Agency (research core Funding No. P1-0035 and J1-3034 and N1-0360).

\end{document}